\documentstyle[12pt,aasms]{article}

\newcommand{\al}{$\alpha$}
\newcommand{\lam}{$\lambda$}

\begin{document}

\title{Spectroscopy of  PKS 0528-250: New Limits on CO Absorption and Emission\footnote{Observations here were obtained with the
MMT, a joint facility of the university of Arizona and the Smithsonian Institution; and the NRAO 12 Meter Telescope. The NRAO is a facility of the
National Science Foundation operated under cooperative agreement by Associated 
Universities, Inc.}}

\author{Jian Ge, Jill Bechtold,  Constance Walker, John H. Black\footnote{Current address:  Onsala Space Observatory, Chalmers University
of Technology,  S-439 92 Onsala, Sweden}}
\affil{Steward Observatory, University of Arizona, Tucson, AZ 85721}

\begin{abstract}
We have obtained a moderate resolution spectrum of the quasar PKS 0528-250 
with the Red Channel Spectrograph on the Multiple Mirror Telescope (MMT)
in order to study a damped Ly$\alpha$ 
 absorption line system at z = 2.8115.  
 We  obtain a new  upper limit for the CO column density for the z = 2.8108 velocity component in the  
z = 2.8115 damped Ly$\alpha$ system. The ionization  of different species 
in this component 
rules out a quasar spectral energy distribution (SED) as  the ionization field,
and implies an   ultraviolet radiation field intensity 
a few times that of the Milky Way value. The estimated total number density
is n(H) $\sim$ 20 cm$^{-3}$.  The physical size for the z = 2.8108 
component implied by these  models is about 40 parsecs. The ionization of different species also suggests a structure with a hot
intercloud medium associated with a H I cloud in this component,
that is, most low ionized ions are from the cold medium where photoionization and
photodissociation dominates.
The highly ionized species may be  from the intercloud medium
 where collisional ionization dominates.
 We also present  newly
 identified Ni II 
absorption lines in the
z = 2.1408 and z = 2.8115 damped Ly$\alpha$ systems. The derived  depletion of 
nickel by dust 
confirms previous results that the dust-to-gas ratio in these two
 damped Ly$\alpha$ systems is 
about 10\% of the Milky Way ratio.   Millimeter  wavelength observations
obtained at the NRAO 12 meter telescope
provide new upper limits on CO (3-2) emission  in the z = 2.8115 damped Ly$\alpha$ system.

\end{abstract}


\section{Introduction}
Damped Ly$\alpha$ quasar absorption line systems may be the high redshift 
analogs of present-day galactic disks (Wolfe {\em et al.}   1986). They dominate
the cosmological mass density of neutral gas in the Universe and hence trace 
the bulk of material available for star formation at high redshifts (Lanzetta
1993). If the physical conditions in the damped Ly$\alpha$ systems are similar to those
in Galactic interstellar clouds, the damped Ly$\alpha$ systems should  show 
absorption from low-ionization  species like C I, trace elements like
Ni, Zn and  Cr and molecules like H$_2$ and CO. Most of these 
species have been detected in absorption  in 
 the damped Ly$\alpha$ systems (e.g. Foltz {\em et al.}   1988;
Meyer {\em et al.}   1986, 1987, 1989; Pettini {\em et al.}   1994, 1996; Ge, 
Bechtold \&  Black 1997; Ge \& Bechtold, 1997), with the notable exception 
of  CO.  Detection of molecules at high redshift would provide a strong 
diagnostic  for the conditions in the clouds that harbor them.

To detect absorption lines from C I, Ni II, Zn II, Cr II and CO, high
signal-to-noise ratio (S/N) spectra are required because
the expected lines are very weak. PKS 0528-250 is one of the brightest quasars known, 
m(1450\AA)=17.73 (Sargent {\em et al.}   1989), and so
is a good object for detailed work. Toward PKS 0528-250,  there are two damped
Ly$\alpha$ systems at z = 2.1408 and z = 2.8115. The z = 2.8115
damped Ly$\alpha$ system  has N(H I) $=2.24\times 10^{21}$ cm$^{-2}$ and  is one of the
highest column density damped Ly$\alpha$ systems known (M{\o}ller \& Warren 1993).

The z = 2.8115 damped Ly$\alpha$ system is a good candidate to detect C I and CO.   Foltz {\em et al.} (1988) have detected H$_2$ absorption 
associated with it.
 Lanzetta (1994, private communication)  and 
Songaila \& Cowie (1995) have further 
confirmed this detection. The redshift for the H$_2$ absorption is z = 2.8118
(Lanzetta 1994, private communication).
 The column density of molecular hydrogen, N(H$_2$) =  2.6$\times 10^{16}$ cm$^{-2}$, a factor of 10  larger than the upper limits for other  damped 
systems (Songaila \& Cowie 1995; Levshakov {\em et al.}   1992 and references therein).
 The  molecular fraction
f=2N(H$_2$)/[2N(H$_2$)+N(H I)]=1.2$\times 10^{-5}$ is 
 similar to the Milky Way diffuse clouds (Savage {\em et al.}   1977). 
 Previous observations show that the 
H$_2$, CO and C I column densities are 
correlated with each other in the Milky Way diffuse clouds, 
e.g. N(H$_2$)/N(C I) is $\sim 10^2$ - $10^6$ and
N(C I)/N(CO) is  $\sim 10$ (e.g. Federman {\em et al.}   1980). Thus, we expect that
the C I and CO column 
densities  are   high enough to be detectable.

M{\o}ller \& Warren (1993)  detected Ly$\alpha$ emission at z = 2.8121.
  The detected Ly$\alpha$ line luminosity of
1.1$\times 10^{42} h^{-2}$ ergs s$^{-1}$  corresponds to a star formation rate (SFR)
of $1.0 h^{-2}$ M$_\odot$yr$^{-1}$, if there is no attenuation by dust  
(we adopt q$_0=0.5$, 
H$_0=100 h$ km~s$^{-1}$Mpc$^{-1}$). The minimum size
of the absorber is about 18$h^{-1}$ kpc.

To study the nature of the z = 2.8115  damped Ly$\alpha$ system we have conducted 
two kinds of observations: (1) high signal-to-noise optical
spectroscopy covering the strongest absorption bands of the CO A-X system 
(0-0, 1-0, 2-0, 3-0, 4-0, 5-0, 6-0), C I $\lambda$ 1560 and some
 Ni II lines; and (2) millimeter observations of CO (J=3-2) emission at 90 GHz.
The observations are described  in  section 2.   Identifications of absorption 
lines  are discussed in section 3.  We discuss our results in
section 4 and
give a summary in the
final section.

\section{Observations}
\subsection{Optical Spectroscopy}
The observations of PKS 0528-250 were obtained on February 14, 1994 with the 
Loral 800$\times$1200 CCD attached to the Red Channel Spectrograph on the MMT 
(Schmidt {\em et al.}   1989).
The 1200 l/mm grating was used in first order with 1$''\times 180''$ slit,
which resulted in  spectral resolution of 2.0 \AA\ (FWHM),
 measured from the comparison lamp lines. The seeing was  0.5$''$ (FWHM).
 We took  three  40 minute exposures with wavelength
coverage from 5200 \AA\ to 6100 \AA . The quasar was stepped by a few arcseconds
along the slit between each exposure to smooth out any residual irregularities 
in  the detector response which remained after 
flat-fielding. An exposure of a He-Ne-Ar-Cd-Hg lamp and a quartz lamp were done
before and after each exposure of the object to provide an accurate wavelength
reference, a measure of the instrumental resolution, and a flat-field correction.
Bias frames were also obtained.
The spectra were reduced in the  standard way using IRAF. 
The summed spectrum weighted by the signal-to-noise 
 from the three exposures is 
shown in Figure 1. The  signal-to-noise ratio
 is about 50 in most of the wavelength coverage.

The continuum was fitted iteratively  using a  
cubic spline routine.  
First the spectrum was divided into some moderately wide bins (40 pixels) and
averaged in each bin.  Then a cubic spline was fit to these averaged points. 
Points which deviated negatively by more than two standard deviations were
flagged  as possible significant lines and rejected for subsequent iterations 
of the fit.  This process was repeated about six times until the fit did not change,
i.e. until the number of flagged pixels above the fit was about equal to
the number of pixels below the fit and the reduced chi-squared of all 
unflagged pixels was near unity (Bechtold 1994). 

Significant lines were found by measuring the equivalent width in a bin equal
to 2.46 times the number of pixels of the FWHM of the comparison lines, and
its error (c.f. Young, Sargent  \& Boksenberg 1982). All such bins where the 
equivalent width was significant at greater than or equal to 3 $\sigma$ 
were flagged. The final equivalent width of each absorption line  was 
measured by specifying the wavelength interval for each feature by 
hand. The central wavelengths, equivalent widths and their errors are listed 
in Table 1. The central wavelength of each line in  Table 1 is the centroid, weighted 
by the depth of each pixel in the line profile below the continuum 
(c.f. Bechtold 1994). The 
error for the central wavelength shown in this table is from the uncertainty 
in the measurement of the equivalent width. We neglected the uncertainty from the 
wavelength calibration because 
the error from the wavelength calibration is very small, typically
 less than 0.02 
\AA .
 All
reported wavelengths are vacuum and have been corrected to the heliocentric
frame.

\subsection{Radio Observations}
Our millimeter observations of the z = 2.8108 damped Ly$\alpha$ system toward PKS 0528-250 were
made in January  1994 at the NRAO 12 Meter telescope.
All of the observations were done with the
 dual polarization SIS 3 mm mixer receiver, 
which provided a  system temperature of 260 K at 90 GHz. The central 
frequency  was set to 
90.74093 GHz 
corresponding to the
heliocentric redshifted CO (J=3-2) emission frequency of the z = 2.8108 damped Ly$\alpha$ system.
The telescope 
half-power beam width (HPBW), $\theta_B$, was 68$''$.
 The frequency calibration  was checked with 
observations of Orion A. The pointing  was checked with observations
of Q 0607-157. 

The spectra were obtained using absolute position switching between the source
position and a reference position 3$^\prime$ away in azimuth. We used a 256
channel filter bank spectrometer with a spectral resolution of 2MHz. We obtained
498 minutes of integration on source in four nights; scans were six minutes in 
length. Each scan had the same three bad channels removed before the scans were
summed. Two noisy scans were omitted from the final summation. Before taking
the mean of a channel, we eliminated channels with more than a 10$\sigma$
deviation from a robust median. At that point, a final  1$\sigma$ deviation of
each channel was calculated (Walker {\em et al.}   1996). The final spectrum is shown 
in Figure 2, where a single linear baseline has been fitted (by iteratively throwing
out channels with intensities greater than 3.5$\sigma$) and then removed. In 
Figure 2, the 1$\sigma$ error is represented as a dashed line. The results are
discussed in section 4.4.

\section{Notes on Individual Systems} 

Table 1   lists identifications of absorption lines in the optical spectrum of 
0528-250.  The redshifts of absorption lines are consistent with those 
identified by previous authors (Morton {\em et al.}   1980; Meyer \& York 1987; Sargent
{\em et al.}   1989). A few weak lines in Table 1 have not been identified and may be spurious.
The z = 2.671 and z = 2.673 C IV absorption line systems are newly 
identified by us.  We describe the
different redshift systems separately in the following.

z$_{ab}=0.9442$ --- The redshift agreement is good between the two components
of the Mg II doublet. The equivalent widths for these two lines are about 
half the values obtained by Sargent {\em et al.}   (1989). The discrepancy may be
caused by  line blending  in their data.  

z$_{ab}=2.1408$ --- The equivalent width of Al II $\lambda$ 1671 from this damped Ly$\alpha$ system
is consistent with that measured by other authors (Morton {\em et al.}   1980; Sargent {\em et al.}   
1989). We identify the absorption line at $\lambda = 5849.26$ \AA\ 
as Al III $\lambda$ 1863. The other member of the doublet,  Al III $\lambda$ 1860 at 
$\lambda=5825.29$ \AA,
is blended 
with the strong absorption line Si II $\lambda$ 1526
from the z$_{ab}=2.8115$ damped Ly$\alpha$ system.    We have confirmed the identifications of
two weak   absorption lines of Ni II $\lambda$ 1709, Ni II $\lambda$ 1741  by Meyer {\em et al.}   
(1989) and  we identify another weak  Ni II transition,
Ni II $\lambda$ 1751. We discuss these further in section 4.2.

z$_{ab}=2.5381$ --- The equivalent widths of the C IV doublet are statistically
consistent with 
those obtained by other authors (Morton {\em et al.}   1980; Sargent {\em et al.}   1989).
We have not detected absorption from Si II $\lambda$ 1526  in our spectrum. 

z$_{ab}=2.6259$ ---This C IV absorption line system is tentatively identified
here. There is a Ly$\alpha$ absorption line 
at $\lambda =4406.57 $\AA\
with W$_{obs} =3.60$\AA\, corresponding to z$_{ab} = 2.6248$ 
(Morton {\em et al.}   1980). However, it is blended with the Si IV $\lambda$ 1402 line 
of the  z$_{ab} = 2.1408$ damped Ly$\alpha$ system. The absorption lines at 5614.02 and 5622.57
\AA\ could also be the absorption lines of Co II $\lambda\lambda$ 1472, 1475
from the  z = 2.8115 damped Ly$\alpha$ system.

z$_{ab}=2.6710, 2.6732$ --- These two C IV absorption line systems are first
identified here. The C IV 1550 absorption line from the z = 2.6710 system 
is blended
with the C IV 1550 line from the z = 2.6732 system.
 The z = 2.6732  system has a moderate Ly$\alpha$ absorption
line in the Ly$\alpha$ forest 
at $\lambda = 4465.42$ \AA\ with 
W$_{obs} =4.84 $\AA\, corresponding to z$_{ab}=2.6732$ 
(Morton {\em et al.}   1980). 

z$_{ab}=2.8115$ --- The equivalent widths of the strong absorption lines Si IV 
$\lambda$ 1393, 1402, Si II $\lambda$ 1526, C IV $\lambda$ 1548, 1550 are  consistent with earlier
measurements of Morton {\em et al.}   (1980), Meyer \& York (1987) and Sargent {\em et al.}   
(1989). We have confirmed that the z$_{ab}=2.8115$ system consists of at least two    
components with z$_{ab}=2.8108$ (A$_1$) and z$_{ab}=2.8132$ (A$_2$)
 respectively (Morton {\em et al.}   1980), corresponding
to a velocity separation of 45 km s$^{-1}$. We
have detected three Ni II absorption lines from the z$_{ab}=2.8115$ system at 
better
than the 3$\sigma$ level. The absorption line at $\lambda = 5222.84 $
\AA\ is identified
as Ni II $\lambda$ 1370.13   at z = 2.8119 
instead of C IV $\lambda$ 1548.20 at z = 2.3733 by Chen \&
Morton (1984) because no absorption line  was detected for
C IV $\lambda$ 1550.77 at the same redshift, and the equivalent width should be 
at least  0.32 \AA\ if both members of the doublet are unsaturated, 
compared to our 5$\sigma$ upper limit of 0.26 \AA. 
 The Ni II $\lambda$ 1370 and Ni II $\lambda$ 1454 lines  exhibit the same 
two redshift components as the low ionized ions  Si II, Fe II, OI 
and Al II lines. The absorption line of Ni II $\lambda$ 1467, which is weaker than Ni II $\lambda$ 1370,
1454, does not exhibit two velocity components 
probably due to limited S/N and resolution. The lines of the  ionized species,
 C IV $\lambda$ 1548, 1550 and Si IV $\lambda$ 1393, 1402 
are seen from the z$_{ab}=2.8108$ system 
  only (Morton {\em et al.}   1980; Meyer \& York 1987).

There is an absorption line at the wavelength corresponding to
C I $\lambda$ 1560.31 
from the z = 2.8115 damped Ly$\alpha$ system
at about the 4$\sigma$ significance level. The rest frame equivalent width of 
this line is $38\pm 10$ m\AA\ which is lower than the 2$\sigma$ upper limit of 50 m\AA\  given 
by Meyer {\em et al.}   (1987). 
If  this is  the  C I absorption line,  then the column density of C I is $2.2\pm 0.6\times 10^{13}$ cm$^{-2}$, assuming that the line is on the linear 
part of the curve-of-growth. 
However,  much higher  quality spectrum with the resolution of about 8  
kms$^{-1}$  and S/N $\approx$ 50 from the KECK HIRES 
by Songaila  {\em et al.} (1995, private  communication) rules out this  
identification. Their spectrum does not show  any absorption lines from C I $\lambda$ 1277.24  and C I 
$\lambda$ 1328.83 to the 3 $\sigma$ upper limit of  N(C I) $\le 6\pm 
2\times 10^{12}$ cm$^{-2}$, which  is much smaller than that from our data. 
 The 4  $\sigma$ absorption  feature
 in our MMT spectrum is probably a statistical artifact. Therefore, 
in what follows, we  adopt the measurements by Songaila {\em et al.}

We have not detected absorption of
the excited  Si II* $\lambda$ 1533 from the A$_1$ component in the z = 2.8115 
damped Ly$\alpha$ system. The 3 $\sigma$ upper limit of equivalent width is 0.118 \AA\, 
which will be used to derive the number density of hydrogen in this
 component (see
section 4.1).
   The absorption of Si II* $\lambda$ 1533 from the A$_2$
component in the z = 2.8115 damped 
system is blended with the Al III $\lambda$ 1863 line 
from the z = 2.4108 damped Ly$\alpha$ system.

The absorption line at $\lambda = 5519.40$ \AA\ is consistent 
with the CO A-X 1447.36 absorption line  at z$_{ab}=2.8132$. While CO $\lambda$
1447.36 is one
of the strongest CO absorption lines (e.g. Field {\em et al.}   1983; Sheffer {\em et al.}   1992), 
 we 
failed to detect the CO A-X $\lambda\lambda$ 1477.52, 1509.70  absorption 
lines  which have similar strengths. The absorption 
line at 5519.40 \AA\ could also be the 
Co II $\lambda$ 1448  line from the A$_1$ component of
 the z = 2.8115 damped 
system. Then the absorption line at  6000.53 \AA\ could be 
Co II $\lambda$ 1574
lines from the same component. However, we also failed to
detect  Co II $\lambda\lambda$ 1466, 1480 in our
spectrum which have similar strengths (Morton 1991). Therefore, the identification for the absorption at $\lambda$ = 
5519.40 \AA\ is uncertain.

\section{Results and Discussion}

\subsection{Ionization}

We have constructed
 models using the CLOUDY program (Ferland 1993)
in order to investigate the ionization  of the observed species, the
spectrum  of the UV ionizing radiation field,
 and the molecular fraction  of the damped Ly$\alpha$
absorption line system at z = 2.8115.

The column densities for the species observed  in the
A$_1$ and A$_2$ components (z = 2.8108, z = 2.8132) of the z = 2.81 damped Ly$\alpha$ system are 
shown in Table 2. Because the resolution of the available spectra is
insufficient to resolve the profiles of the lines listed in Table 2, we use 
a curve-of-growth analysis  to infer column densities.
 We adopt the Doppler
parameter b = 60 km s$^{-1}$ for  lines from the A$_1$ component 
and b = 30 
km s$^{-1}$ for lines from the A$_2$ component which are obtained 
from the curve-of-growth of Fe II (Meyer {\em et al.}   1989). If the absorption lines
from the A$_1$ and A$_2$ components
are blended with each other, 
we adopt b = 100 km s$^{-1}$ 
obtained from fitting the  Lyman lines (Foltz {\em et al.}   1988). However, the Si IV and C IV doublets in the A$_1$ component
 suggest that the Doppler parameter b = 60 km s$^{-1}$  derived for
Fe II lines does not apply. A higher Doppler
parameter b = 80 km s$^{-1}$ gives a better fit. 
Therefore, we adopt b = 80 km s$^{-1}$ to calculate
 column densities of  high ionization species, C  IV, Si III, Si IV, N V, O VI in the A$_1$
component (Table 2). 
The b-values  indicate there are probably several velocity components  blended
with each other in each line. However,
the uncertainties in the column densities from single b-value curve-of-growth 
analysis may  be less than a factor of 2  (Jenkins  1986). 
In fact, our derived column densities are consistent with the
 results given by Songaila \& Cowie (1995), and Lu  {\em et al.} (1996) based on
higher resolution spectra.

 As the input to CLOUDY, we adopt a metallicity of 
10\% of the  solar value for all the elements in  both A$_1$ and A$_2$
components of the z = 2.8115 damped Ly$\alpha$ system. We also consider the relative 
depletions by dust grains to Zn (about 0.6-0.7 in the logarithm unit) of 
heavily depleted elements such as
Al, Si, Ca, Fe and Ni (Meyer {\em et al.}   1989).  The dust-to-gas ratio is  
 10\% of the Milky Way and the constituents of dust grains are assumed  the
same as the Milky Way. The total neutral hydrogen column density N(H I)
for the two components is $2.24\times 10^{21}$ 
cm$^{-2}$ (M{\o}ller \& Warren 1993).
 It is obtained by fitting the profile of the damped
Ly$\alpha$ absorption line including  the Ly$\alpha$ emission 
of the quasar because the redshift of the damped Ly$\alpha$ system (z = 2.8115) is about
the same as that of the quasar (z = 2.765, Morton {\em et al.}   1980; M{\o}ller \& Warren 1993). The neutral hydrogen column density 
for each of the two components in 
the z = 2.8115 damped Ly$\alpha$ system is  uncertain. However,  
similar absorption of dominant ions of most species, such as O I, Si II, S II,
Fe II, Ni II,  in these two components 
suggest that  the total neutral hydrogen column density in each component is
also 
similar (cf. Table 2).  We take the total
column density as the approximate value for each of the components in the
following discussion. There is therefore at least  a factor of 2 uncertainty 
in the 
results from CLOUDY from the  uncertainty in the neutral hydrogen column density
in each velocity component.

 Since  the damped Ly$\alpha$ system is possibly
near the quasar, the UV flux from the quasar can affect the 
ionization. We therefore
have considered  five types of  spectral energy distributions (SED). Figure 3 shows these SEDs, along with the ionization potential of several ions.

\subsubsection {Quasar  SEDs} 

 We use a
power-law  with index $\alpha =  1.0$ for 
$\lambda \ge 912 $\AA , where $f_\nu
\propto \nu^{-\alpha}$, measured for PKS 0528$-$250  by Sargent {\em et al.}   (1989).
 We  extrapolate the power to longer
wavelengths which has little affect on our results. 
The quasar energy 
distribution at wavelengths shorter than the Lyman limit is  uncertain. 
We therefore consider two different quasar SEDs: a(1) and a(2) (Figure 3).
 In these two models, the 
upper limit to the X-ray flux from the Einstein IPC at 1.0 keV is adopted for 
the wavelength longer than 0.2 keV (Wilkes {\em et al.}   1994). In a(1),
we assume a power-law with  index between the Lyman limit and 0.2 keV of
  $\alpha = 1.0$; this is the hardest spectrum that seems reasonable. In a(2), the SED between 
 the Lyman limit and 0.2 keV is a power law  connecting  these two points.

\subsubsection{1 Gyr old constant SFR galaxy SED}

We consider the models for the evolution of galactic SED's, as calculated by Bruzual \& Charlot (1993).
The shape of the UV SED in the  $\lambda \la 3000$\AA\ region does not change
after about 0.1 Gyr if the SFR is constant, so  we  consider the SED of 1 Gyr old galaxy as representative. This is shown in the Figure 3 as model b.
For an age less than 0.1 Gyr the SED is similar to the SED
of an instantaneous starburst galaxy, which we  consider  in the next model.

\subsubsection{0.001 Gyr old instantaneous starburst galaxy SED}

 Compared with a 1 Gyr old constant SFR galaxy SED model, this model lacks  hard UV photons because of the lack of AGB stars. 
This is shown in Figure 3 as model c.

\subsubsection{Milky Way SED}
We adopt the parameterization of the Milky Way radiation field by Black (1987). This includes starlight, cosmic background radiation, thermal 
emission of dust, hot interstellar gas heated by supernova remnants,
 stellar winds, and UV radiation from extragalactic sources such as QSOs and nearby
active galaxies. The energy distribution  between 13.6 eV and  54 eV is a 
power law extrapolated 
from other wavelengths. This is shown in Figure 3 as model d. 

\subsubsection{Power-law SED  with spectral index $\alpha$ = 2.7}
The SED is a single power-law from radio to X-ray wavelengths. The energy
distribution between the Lyman limit and 1100 \AA\ is harder than the 
Milky Way SED,  the constant SFR galaxy SED and the starburst 
galaxy SED, but is much softer than the quasar SEDs. This is shown in 
Figure 3 as model e.

\vspace{0.4 in}

 Figure 4 shows the CLOUDY results for low ionization species. 
The ordinate is the column density of the various species, and the 
abscissa is the ionization parameter, U = $\phi$(H)/n(H)c, where $\phi$(H)
is the surface flux of hydrogen-ionizing photons, n(H) is the total hydrogen
number density (ionized, neutral, and molecular hydrogen),
 and c is the speed of light.
  CLOUDY shows that the column densities of
 O I, S II, Fe II, Si II and Ni II, which are the dominant 
ions of those species in the H I
dominant regions, are not sensitive to different SED models over a large range of ionization parameter. Since  all SED models 
 give similar results for these ions,  
we only show the result from model e in
Figure 4. The abundance of these ions from  CLOUDY are consistent with
the observed ones within about a factor of two indicating
 that we have a reasonable set of
assumptions for the abundance and depletions.

Figure 5 shows the calculated  and observed ratios of N(H$_2$)/N(C I) 
in the A$_1$ component of z = 2.8115 damped 
system. The common trend in the figure of N(H$_2$)/N(C I) vs. N(C I) can be 
explained by molecular hydrogen chemistry.  
The main destruction process of H$_2$,  photodissociation  through the 
process of spontaneous radiative dissociation by the Lyman and Werner bands,
is initiated by discrete line absorptions. The lines are very narrow. Thus,
the lines rapidly become optically thick, and then H$_2$ shields itself 
effectively, so that most of the hydrogen forms H$_2$ (e.g. van Dishoeck 1990).
 This is why we see a rapid increase of N(H$_2$)/N(C I) vs.
N(C I) when the UV flux drops below a certain threshold. After that,
the abundance of C I still increases as the UV decreases, but the 
abundance of H$_2$ 
increases slowly due to lack of formation material, hydrogen, which has 
already been changed into H$_2$. This is why 
 the ratio of N(H$_2$)/N(C I) decreases after the rapid increase phase.
The best fit to the observed ratios is the starburst  model (b) and (c). 
Model (d)  and (e) can fit the observed ratios within
about 3$\sigma$.  
The quasar  SED models (model a(1) and a(2)) 
 cannot explain the observed ratios.  The observed ratios are best explained by
  the soft UV energy distribution models, which suggest 
 stars contribute most of 
  the ionizing radiation field 
in the A$_1$ component of z = 2.8115 damped Ly$\alpha$ system. 
  Thus, the SED model where the UV flux is mainly contributed by
 PKS 0528-250 can
be ruled out, i.e. the distance
between the A$_1$ component and the quasar 
is probably  large enough ($\ga$ 1 Mpc) that the UV  flux from 
the quasar does not dominate the UV background radiation field in the A$_1$ component.

 Figure 6 shows the CLOUDY results for the high  ionization
species assuming the  total neutral hydrogen column density used in Figure 4. 
None of the model SEDs can explain the C IV, N V, Si IV, O VI and Si III 
column densities.
Hence the absorption lines for 
these high ionization species are not coming from the same material as the 
absorption for the low 
ionization species. However, assuming the equilibrium between collisional ionization and radiative and
dielectronic recombination, we find that T $\sim 1\times 10^5$ K can 
 explain the relative ratios
of these species (Shull \& Van Steenberg 1982).
The A$_1$ component may be a combination of a disk-like
system and a halo, or   warm and cold H I clouds
 associated with a hot intercloud medium.

Using the CLOUDY model, we can further estimate 
 the average value of  the  UV radiation field
 along the sightline  to the quasar in the A$_1$ component.
In order to calculate this, we  need to know the number density of neutral 
hydrogen. 
We estimate this density using the relative population ratio of the 
excited state Si II$^*$ $\lambda$ 1533 \AA\ to the ground state Si II $\lambda$ 1526 \AA. 
In the H I dominant region with density less than $\sim 10^4$ cm$^{-3}$, this
population ratio  can be expressed 
as
\begin{equation}
 \frac{N(Si~II^*)}{N(Si~II)} =\frac {n_e<\sigma v>_e + n_p<\sigma v>_p + n_H <\sigma v>_H + n_{H_2}<\sigma v>_{H_2}}{A_{10}}, 
\end{equation}
 where $n_e$, $n_p$, $n_H$ and $n_{H_2}$ are, respectively, 
the electron, proton, hydrogen-atom and molecular hydrogen
 number densities, $<\sigma v>$ is the collision rate, and 
$A_{10}$ is the spontaneous transition probability (Bahcall \& Wolf 1968). 
 CLOUDY shows that the 
relative electron density $n_e/n_H$ is about $\sim 10^{-2}$ at the  outside to
$\sim 10^{-4}$ at the center of the cloud, and the equilibrium temperature is from $\sim 
10^3$ K in the  outside to $\sim 100$ K in the  center.  We therefore
 can roughly neglect the contributions of the fine structure
excitation   by the proton and molecular hydrogen collisions (Bahcall
\& Wolf 1968). Then, 
\begin{equation}
 \frac{N(Si~II^*)}{N(Si~II)}\approx \frac {n_H <\sigma v>_H+n_e<\sigma v>_e}{A_{10}}, 
\end{equation}
 where $<\sigma v>_H = 1.3\times 10^{-9}~exp(-413/T)$ cm$^3 s^{-1}$ 
(Bahcall \& Wolf 1968), and
$<\sigma v>_e = 3.8\times 10^{-7}~exp(-413/T)$ cm$^{-3}$ s$^{-1}$ (Dufton \& Kingston 1991). Taking the observed ratio $N(Si~II^*)/N(Si~II)< 1.41\times 10^{-3}$, T $\sim 10^3$ K and $n_e/n_H \sim 10^{-2}$, we get $n_H < 47$ 
cm$^{-3}$. 
 CLOUDY gives a best fit  log U $\approx -$2.9 and $-$2.6 for the model b and
c, respectively.
Thus, the upper limit of the neutral
 hydrogen density implies
an upper limit of 1-2$\times 10^9$ cm$^{-2}$ s$^{-1}$ for the hydrogen-ionizing photon flux  in the A$_1$ component. This upper limit is
 about  100 times
larger than the Milky Way value of $\sim 1\times 10^7$ cm$^{-2}$s$^{-1}$ 
(Black 1987).  Because Si II$^*$ traces warm phase regions 
 outside of the cloud (cf. Morton 1975), the above result is more 
suitable for outside region of the cloud. For the central part of the absorber
cloud, the population ratio of the  H$_2$ J = 4 to J = 0 levels, N(J=4)/N(J=0)
= 4.5$\times 10^{-3}$ (Songaila \& Cowie, 1995), provides
 $Rn \approx 6\times  10^{-17}$  s$^{-1}$, where  $R$ is the H$_2$ 
formation rate, n = n(H) + 2n(H$_2$) (see Jura 1975 for details).
 We scaled the H$_2$ formation rate to R $\sim 3\times 10^{-18} $ 
cm$^{3}$s$^{-1}$, assuming the  dust-to-gas ratio is 10\% that of the Milky Way.
 The  density is then $<n> \approx <n_H> \sim$ 20
cm$^{-3}$.    We therefore can use this  value and
other measurements to 
estimate the physical properties in the central part of the cloud and compare 
them with
the CLOUDY results. In the   central cloud where the kinetic temperature
is around 100 K (Songaila \& Cowie 1995), 
C II provides most of the   electrons 
(e.g. Morton 1975), so 
\begin{equation}
\frac{N(C II)}{N(H I)}\approx\frac{n(e)}{n(H)}\approx 7.6\times 10^{-5},
\end{equation}
 where we have assumed a spatially homogeneous distribution of C II
ions, electrons and 
hydrogen atoms. Thus,  $n(e)\sim 10^{-3}$ cm$^{-3}$.
The equation for C I photoionization equilibrium can be expressed as
\begin{equation}
n(e)=\frac{N(C I)}{N(C II)} \frac{\Gamma(I)}{\alpha(T)},
\end{equation}
where we have also assumed a homogeneous distribution of C I; 
the photoionization 
rate, $\Gamma(I)$ (s$^{-1}$),  is  a function of the radiation field intensity,
 I; and $\alpha(T)$ (cm$^3$ s$^{-1}$) is the radiative recombination rate
 coefficient, which is a function of kinetic temperature. The kinetic 
temperature in the H$_2$ containing cloud is between 74 K and 270 K, where 
74 K is derived from the population ratio of the H$_2$ in the first excited 
rotational state to the ground rotational state 
and 270 K is derived from the b = 1.5 km s$^{-1}$
 for the H$_2$ absorption lines (Songaila \& Cowie 1995). Therefore we choose 
$\alpha(T) = 1.0\times 10^{-11}$ cm$^{3}$s$^{-1}$ (e.g. Snow 1977). 
From the measurements, we derive N(C I)/N(C II)$\lesssim 3.5
\times 10^{-5}$. Putting these together, we get $\Gamma(I)\gtrsim
2.9\times 10^{-10}$ s$^{-1}$ which is  a factor of few larger than 
typical value 
of the $\Gamma(I)\sim 10^{-10}$ s$^{-1}$ in the Milky Way ISM 
(de Boer {\em et al.}   1973). This value is
consistent with what we expect for the outer region of the cloud. Thus, 
all results 
derived here are consistent with those derived
 from the CLOUDY analysis. 

We can further
use these estimates of the physical parameters to estimate  the  size 
of the z = 2.8108 cloud. The averaged value for the electron density is 
derived to be $<n(e)>\sim 10^{-3}$ cm$^{-3}$,
N(C II)$\approx$ N(e)$\sim <n(e)>l$, thus, the physical size for the 
z = 2.8108 cloud, is $l \sim 40$ pc. 

For the A$_2$ component, the lack  
of highly ionized ions may mean that the A$_2$ component is not 
associated with any halo gas. The photoionization models described above cannot explain the lack of H$_2$ and C I, along with the possible
detection of CO, and  special circumstances
 for molecule formation may apply.

In summary, we can account for  the z = 2.8108 absorption line 
system (A$_1$) containing  both  hot and cold, neutral gas. 
The cloud  medium contributes most
 absorption lines  of low  ionization
  species, while the hot  medium contributes
most absorption lines of highly ionized  species. This structure is similar to 
that of  the  interstellar diffuse clouds in the Milky Way.  The UV flux intensity in 
this absorber is  a few times that of the
typical value in the Milky Way ISM.

\subsection{Dust Depletion  of the  Damped Ly\al\ Systems}

Pettini {\em et al.}   (1994) have surveyed about a dozen damped Ly$\alpha$ systems to measure the
abundance of Zn and Cr and  have claimed that the metallicity
of the damped Ly$\alpha$ systems at z $\sim$ 2 is about 1/10 of the solar value. They used the 
relative abundance [Cr/Zn] to deduce the dust-to-gas ratio in the damped 
systems, which is about 1/10 of the Milky Way. This result is  consistent 
with that from the dust  reddening measurements of quasar spectrum slope
by Pei {\em et al.} (1991). However,  Lu {\em et  al.} (1996)
recently claimed that the  overabundance of Zn  relative to Cr may be 
intrinsic to the stellar  nucleosynthesis  in these absorbers  instead of 
dust depletion (however, see Pettini {\em et al.}  1996; Smith  {\em et al.} 
 1996). Here, we assume
that the underabundance of Cr and Ni  relative to Zn is caused by dust 
depletion.
Compared with  Cr, Ni is   perhaps a better   element to use to deduce
 depletion.
Like Zn and Cr, Ni is an iron group element and  it traces Fe to about 
[Fe/H] $\sim -2.5$ or even lower (Wheeler {\em et al.}   1989; Ryan {\em et al.}   1991). Ni 
is also  more heavily depleted
 than Cr in the Milky Way (Jenkins, 1987):  0.35\% Ni remains in the gas phase compared
to 0.62\% for Cr in the diffuse cloud toward $\xi$ Per  (Cardelli {\em et al.}   1991).
Because the ionization potential of Ni I is 7.635 eV which is smaller than that of H I, 
but  the ionization potential of Ni II is 18.168 eV which is larger than that of H I, 
Ni II is  the dominant 
species of Ni in the neutral hydrogen dominant region. Consequently, the ratio of
N(Ni II)/N(H I) can reflect that of  Ni/H gas
phase abundance without the need to account for unobserved ion stages. For 
example, under the condition that best fits the ionization structure of the
A$_1$ component in the z = 2.8115 damped Ly$\alpha$ system, model (b),
log U = $- 2.9$ and n(H)=1 cm$^{-3}$, N(Ni II)/N(Ni) = 97.7\%.
  Moreover, 
there are more UV absorption lines of Ni II than  Cr II. If we can observe
more than two weak absorption lines of Ni II, we can use a curve-of-growth analysis
to get a more accurate estimate of the column density. Another  potential 
advantage of using Ni II instead of  Cr II is that the Ni II transitions are observable with
ground-base telescopes until redshifts about 5 without strong sky emission
 because
the strong UV transitions of Ni have rest wavelengths between  1400 and 1700 \AA\
compared to $\sim$ 2000 \AA\ for Cr and Zn.

Our measured 
equivalent widths of Ni II $\lambda$ 1709, Ni II $\lambda$ 1741 and Ni II 
$\lambda$ 1751  in the z$_{ab}=2.1408$ system 
correspond to column densities 3.4$(\pm 1.1)\times 10^{13}$ cm$^{-2}$, 3.8$(\pm 
0.9)\times 
10^{13}$ cm$^{-2}$ and 3.2$(\pm 0.9)\times 10^{13}$ cm$^{-2}$, respectively,
when we adopt a Doppler parameter b = 30 km s$^{-1}$ (Meyer {\em et al.}   1989).
They are consistent with the values obtained by Meyer {\em et al.} 
 The average column density of Ni II is 3.5$(\pm 0.5)
\times 10^{13}$ cm$^{-2}$. The H I column density in this system is 5$(\pm 1)\times 10^{20}$ 
cm$^{-2}$ 
(Morton {\em et al.}   1980). Thus, Ni/H is 7.0$(\pm 0.2)\times 10^{-8}$, implying that
 Ni is depleted by no more than a factor of 27$\pm$4 in the z$_{ab}=2.1408$ 
damped Ly$\alpha$ system with respect to the solar ratio of 1.9$\times 10^{-6}$ (Withbroe
1971) or [Ni/H] $=-1.43\pm 0.06$.
The upper limit of N(Zn II), 2.6$\times 10^{12}$ cm$^{2}$,  measured by Meyer
{\em et al.}   (1989) indicates that the metallicity in the 2.1408 system is $\le$ 10\% of 
solar abundance.
  The measured residual depletion  in the z = 2.1408 damped Ly$\alpha$ system
 is [Ni/Zn]$\ge-0.55$. For comparison, [Ni/Zn] $=-1.92$ in 
the diffuse cloud  toward $\xi$ Per of the Milky Way which has N(H) $=1.97(\pm 
0.35)\times 10^{21}$ cm$^{-2}$ (Cardelli {\em et al.}   1991). This is the typical 
value in the Milky Way (c.f. Jenkins 1987).  The limit for  [Ni/Zn] 
implies that 
dust grains in the z = 2.1408 damped Ly$\alpha$ system contain about 70\% of the  total Ni, while dust in 
our    Galaxy contain about 99\% of the Ni. Thus, the dust-to-gas ratio 
is $\le$7\% of that in the Milky Way. The low dust content  implies an inefficient
molecular hydrogen formation rate, which suggests low total molecular mass. 
Hence, if the measured dust-to-gas ratio  is  typical  of 
the interstellar medium in the z = 2.1408 damped Ly$\alpha$ system,
 our result is
consistent with nondetection of CO emission by  Wiklind \& Combes (1994),
 who obtained an
 upper limit for the total molecular mass of $3.2\times 10^{11} h^{-2}$ M$_\odot$ integrating over 1100 km s$^{-1}$, but not consistent with
the huge amount of molecular mass, M$(H_2) =$7$\times 10^{11} h^{-2}$ M$_\odot$, reported
by Brown \& Vanden Bout (1993).

The column densities for the newly identified Ni II $\lambda$ 1370, Ni II $\lambda$ 1454 and Ni II $\lambda$ 1467 lines
in the z = 2.8115 damped Ly$\alpha$ system are shown in Table 2.
 The average column density
of Ni II is 6.5$(\pm 0.6)\times 10^{13}$ cm$^{-2}$ in the A$_1$
 component and 3.3$(\pm 
0.5)\times 10^{13}$ cm$^{-2}$ in the A$_2$ component. Together, we get 
9.8$(\pm 0.8)\times 10^{13}$ cm$^{-2}$ for the two components 
which is 
consistent with the column density obtained by Meyer {\em et al.}   (1989) obtained by
measuring the equivalent widths of Ni II $\lambda$ 1709 and $\lambda$ 1741.
The H I column density for the 2 velocity components together in the damped  system
is 2.24$(\pm 0.05)\times
10^{21}$ cm$^{-2}$ (M{\o}ller \& Warren 1993). Thus, Ni/H is 4.4$(\pm 0.2)\times 10^{-8}$, so that Ni is depleted by no more than a factor of 43$\pm 2$
with respect to the solar value or [Ni/H] $=-1.63\pm 0.02$. The depletion of
Zn is [Zn/H] $=-0.91$ (Meyer {\em et al.}   1989), so the metallicity
is about  10\% of the solar value. The residual depletion of Ni
relative to Zn,  [Ni/Zn] $=-0.72$ which means about 80\% of the Ni
is contained in dust grains. Consequently, the dust-to-gas ratio is about  
8\% of that of the Galaxy if dust grains cause the  depletion of Ni 
relative  to Zn.

\subsection{CO absorption}
In Figure 7, we show our spectra which cover the  five ultraviolet CO bands
(A-X 0-0, 1-0, 2-0, 3-0, 4-0). We mark the central positions of the CO lines
of the two velocity components (z = 2.8108, 2.8132) in each spectrum. 
The CO 3-0 $\lambda$ 
1447 line exhibits absorption at 5887.47 \AA\ at the expected position
 of z = 2.8132 within the 
 uncertainties. If this identification is correct, the CO column density is  4.7$(\pm 1.6)\times 10^{13}$ cm$^{-2}$ assuming the line is unsaturated.
 However, the absorption line at 5887.47 \AA\ could also be the 
Co II 1448 line at
z = 2.8117. The corresponding column density is $3.9(\pm 1.3) \times 10^{13}$ 
cm$^{-2}$, which is consistent with the upper limit obtained by Songaila \& 
Cowie (1995).

In order to improve the  CO column density limit
 in the A$_1$ component, we constructed a 
composite CO spectra from the 5 CO lines using a  method  similar to that employed 
by  Levshakov {\em et al.}   (1989). To obtain the 
composite spectrum, we first shifted the spectral regions containing the CO absorption 
bands to rest frame 1477.52 \AA\, the wavelength of the 
CO A-X 2-0 band, which has
the largest oscillator strength in the CO A-X system (Field {\em et al.}   1983), then normalized
to the continuum,  and  averaged with each pixel weighted by
the inverse of its variance. In order to avoid 
the possible
effect of telluric absorption at $\lambda$ = 5892.08 \AA\ and z = 2.8108
 C IV absorption
at 5900.16 \AA\, we used the fitted continuum
value instead of the observed flux in our coaddition. The final 
composite spectrum is shown in Figure 7. No absorption of CO from
the A$_1$ component is found. The 3$\sigma$ upper limit to the 
equivalent width in the rest frame
 is 17 m\AA .  Using the weighted mean
of the oscillator strengths of the
considered bands, we derive an upper limit on the CO column density of the A$_1$
component of N(CO)$< 2.9\times 10^{13}$ cm$^{-2}$.  Thus the ratio N(CO)/N(H I)$\le 
1.3\times 10^{-8}$ for the A$_1$ component, provided that the total neutral
hydrogen absorption in the z = 2.8115 damped Ly$\alpha$ system 
is from this component.

\subsection{CO emission}

Figure 2 shows the summed spectrum of CO (J=3$\rightarrow$2) from the damped 
Ly$\alpha$ system z = 2.8115 toward PKS 0528-250. 
We have not detected any emission
from the z = 2.8108 and z = 2.8132 components. The 3$\sigma$ upper limit of the observed integrated line 
intensity I$_{CO}=\int T_{mb} dv \le 3 \sigma_{rms} \Delta v/\sqrt N_{chan} 
= 0.141$ K km s$^{-1}$ for the A$_1$ component, and 0.142 K km s$^{-1}$ for the z = 2.8132
component,  where $\Delta v$ = 100 km s$^{-1}$
 is the integrated velocity interval around the absorption components,  and $N_{chan}$ is the  number of channels in the  velocity interval of 100 km s$^{-1}$.

The CO luminosity for the z = 2.8115 damped Ly$\alpha$ system can be expressed,
\begin{equation}
 L_{CO} =  \Omega_S~D_A^2~\int {T_{b} dv}, 
\end{equation}
 where $T_{b}$ is the brightness temperature of the source, $\Omega_S$ 
is the solid angle subtended by the source, and $D_A$ is the angular size 
distance. 
\begin{equation}
D_A = D_L/(1+z)^2,
\end{equation}
 $D_L$ is the luminosity distance,
\begin{equation}
 D_L =c~H^{-1}_0~q_0^{-2}~\{zq_0 + (q_0 -1) [\sqrt{(2q_0~z + 1)} -1]\}. 
\end{equation}
 The observed temperature of the source $T_{mb}$ is defined as
\begin{equation}
 T_{mb} = \frac{\int\int\limits_{\Omega_S} \frac{T_{b}(\theta, \phi)}{(1+z)}~P(\theta, \phi)~d\Omega}{\int\int\limits_{4\pi} P(\theta, \phi)~d\Omega}, 
\end{equation}
 $P(\theta, \phi)$ is the normalized power pattern, and the factor 
(1+z) is from the expansion of the universe. Moreover, the
 angular size of about 2.5$''$ for  the Ly$\alpha$ emission region in the 
z = 2.8115 damped Ly$\alpha$ system  (M{\o}ller \& Warren 1993) is much smaller than the beam size of the telescope of 68$''$, 
so if the  CO emission region has similar size as the 
Ly$\alpha$ emission region, then T$_{mb}$ can be further expressed as
\begin{equation}
 T_{mb} \approx \frac{T_{b}~\Omega_S}{(1+z)~\Omega_B},
\end{equation}
 where $\Omega_B = \pi(\theta_B)^2/4ln 2$ is the telescope beam solid angle. 
Then the CO
line luminosity can be expressed as
\begin{equation}
 L_{CO} = I_{CO}~\Omega_B~D_L^2~(1+z)^{-3}~~K~km~s^{-1}~pc^2,
\end{equation}
  The molecular mass can be written as
\begin{equation}
M_{H_2} = \alpha~L_{CO},
\end{equation}
 where $\alpha$ is the CO to H$_2$
 conversion factor.   Adopting  the standard 
conversion factor, $\alpha = ~4.8 M_\odot~ 
(K~km~s^{-1}~pc^2)^{-1}$ (Sanders {\em et al.}   1991), we obtain a
 3$\sigma$ upper limit for the total molecular mass of $1.87\times 10^{11}h^{-2}~M_\odot$  in the A$_1$ component, and
1.88$\times 10^{11}h^{-2} M_\odot$ in the A$_2$ component. These 
upper limits are about a factor of two better than previous observations
of  Wiklind \& Combes (1994)
applying the same model assumptions to
their T$_b$. 

However, the adopted standard conversion factor  could introduce some 
uncertainty  in the
total molecular mass derived here.  It can be
 easily shown that  the 
conversion factor $\alpha$ for  gravitationally bound 
(virialized) clouds, $\alpha \propto \sqrt {n(H_2)}/T_b$. Hence, the total 
molecular mass in the z = 2.8115 damped Ly$\alpha$ system 
could be overestimated 
if the clouds in it have a higher 
brightness temperature than the Milky Way clouds, whereas the total mass could be underestimated if the
clouds have higher density. Furthermore, there is a weak
 dependence of $\alpha$ on the CO
abundance, $\alpha \propto [X(CO)]^{-1/4}$ (Radford {\em et al.}   1991). If the CO
abundance in the z = 2.8115 damped Ly$\alpha$ system is considerably lower than the Milky
Way molecular clouds, the abundance effect  will lead to an underestimate of the
total amount of the molecular gas. From these  considerations,  the 
uncertainty in estimating the molecular mass for the z = 2.8115 damped Ly$\alpha$ system is a factor of a few. 

For comparison, the total molecular mass 
in the Milky Way is 2$\times 10^{9}~M_\odot$ (e.g.
Sanders {\em et al.}   1991), the total molecular mass in
the ultraluminous IR galaxy  Arp 220 is
$2\times 10^{10}~M_\odot$ (e.g. Solomon {\em et al.}   1990), and the 
total molecular mass
of the hyperluminous IR galaxy F10214+4724 at z = 2.286, is $1.1\times 10^{10}h^{-2}~M_\odot$ (e.g. Downes et  al. 1995).  Thus, 
the derived upper limit of the molecular mass in the 
z = 2.8115 damped Ly$\alpha$ system  is higher than any ultraluminous
IR galaxy.
On the other hand,  our optical 
observations suggest that this damped Ly$\alpha$ system is an analog of a Milky Way-like 
normal galaxy at high redshift. 
In order to detect  the molecular mass 
in this system 
if  it is similar to the Milky Way, an rms/channel of $\sim$10$^{-2}$ mK is
needed, using the same telescope size, system temperature and bandpass. 
This requires an integration time 10,000 times longer than obtained here.
However,  interferometers could be  used to   reach interesting
detection levels (e.g. Omont {\em  et  
al.} 1996; Ohta {\em et  al} 1996).

\section{Summary}
Our main results are  the following.

 1. The ionization  
of different species in the A$_1$ component of the z = 2.8115 damped Ly$\alpha$ system
 rules out the possibility that 
the quasar UV flux dominates the UV ionizing field in this component,
 i.e. the quasar should be 
more than $\sim 1$ Mpc away from the damped Ly$\alpha$ system.  
The ionization  further suggests that the shape of the
UV background radiation field  in this component is similar to 
the Milky Way. 
The UV field intensity   is  a few 
 times that of  the Milky  Way value which further implies that 
 the SFR in this component 
 is probably similar to  the  SFR of 
the Milky Way if the total mass in this component is similar to the 
Milky Way. This result is consistent with the results from Ly$\alpha$ observations (M{\o}ller \& Warren 1993). The physical size for this component
is about 40 pc.
The ionization of different species  suggests a structure with a hot
intercloud medium associated with a H I cloud in this component,
i.e. most lowly ionized ions are from the cold medium where photoionization and
photodissociation dominates.
Most highly ionized ions are from the intercloud medium
 where collisional ionization dominates.

2. We have not detected 
 CO absorption  in the rest UV spectrum from the A$_1$ component of the 
z = 2.8115 damped system.  The  3$\sigma$ upper limit of CO column density is N(CO)$\le 
2.9 \times 10^{13}$ cm$^{-2}$ for this component. The ratios of the N(CO)/N(H I)
and N(CO)/N(H$_2$) are in the range of the values for the Milky Way diffuse
clouds  (Federman {\em et al.}   1980).  The absorption line at 5519.40 \AA\ could be 
 a  CO absorption line from the A$_2$ component; however, it also could be a 
Co II absorption line  from A$_1$ component. Higher resolution and 
higher signal-to-noise observations are needed to confirm the identification of this 
line.

3. Analysis of photoionization and photodissociation of H$_2$ and C I 
suggests that the ratio N(H$_2$)/N(C I) 
is  a good 
indicator of the shape of the radiation field in the ISM of
 damped Ly\al\ absorbers.

4. Newly identified Ni II absorption lines  show that the dust-to-gas ratios
in the z = 2.1408 and z = 2.8115 damped Ly$\alpha$ systems are about 10\% of that of the
Milky Way, which are consistent with previous results.

5. We have not detected  CO (3-2) mm emission from the z = 2.8115 damped 
Ly$\alpha$ system.
The 3$\sigma$ upper limits for the mass of H$_2$ in 
the A$_1$ and A$_2$ components in the
z = 2.8115 damped system are 
$1.87\times 10^{11}h^{-2}~M_\odot$ and $1.88\times 10^{11}h^{-2}~M_\odot$,
respectively. The results still cannot rule out the possibility that   the z = 2.8115 damped Ly$\alpha$ system is similar in molecular mass
 to ultraluminous IR galaxies.

\acknowledgements 
We are grateful to Dr. A. Songaila Cowie for providing important
comments and  data in advance of
publication. We thank Dr. J. Shields for helpful discussions. We  thank Dr. G. Ferland
for providing  his CLOUDY program. We thank Dr. G. Bruzual and 
 Dr. S. Charlot for providing   their Galaxy Isochrone 
Synthesis Spectral Evolution 
Library (GISSEL). We also wish to thank the
staffs of MMTO and NRAO for all of their help. 
 This research was supported by 
NSF  AST-9058510 and NASA grant NAGW-2201.

\newpage
\normalsize
\centerline{\bf Figure Captions}

Figure 1.---The  spectrum of PKS 0528-250 obtained with the MMT Red Channel 
Spectrograph. The lower curve presents the 1 $\sigma$ error as derived 
from count statistics in the object and night sky spectra. The features  identified at better 3$\sigma$ significance  in Table 1 are indicated by tick marks. 

Figure 2.---The summed CO(3$\rightarrow$2) millimeter wavelength spectrum of PKS 0528-250 observed with NRAO
12 Meter telescope. The intensity scale is main-beam brightness temperature (K)
and the velocity scale is the offset (km s$^{-1}$) from the optical redshift,
z = 2.81086. The two velocity components are marked. A fit to  the baseline has been 
subtracted. The upper line presents the 1 $\sigma$ error derived from all
scans after removing bad scans. 

Figure 3.---Spectral energy distributions adopted for
 photoionization calculations.
 a(1) and a(2) are quasar SEDs. (b).  1 Gyr old constant SFR
galaxy SED. (c). 0.001 Gyr old 
instantaneous starburst galaxy SED. (d). Milky Way
SED. (e).  power-law SED with $\alpha = 2.7$. Ionization potentials
 for several ions are marked.

Figure 4.---Ionization models for the H I dominant region 
with N(H I) = $2.24\times 10^{21}$ cm$^{-2}$. The ordinate is the
column density of various ions, the abscissa is the log of the
ionization parameter U.
The figure  shows results of dominant ions, O I, Fe II, Si II, S II and Ni II
in the model e---power-law SED model with $\alpha = 2.7$. Solid lines are
from CLOUDY and dotted lines are the observed values.  

Figure 5.---  The relative ratios of N(H$_2$)/N(C I) vs. N(C I).
The results are shown for  SED models described in Figure 3. The vertical
lines  show the  upper limit (3$\sigma$) of  N(C I). The horizontal lines show 
the lower limit (3$\sigma$) of the ratio of N(H$_2$)/N(C I).

Figure 6.--- Model predictions for the highly ionized species, O VI, N V,
C IV, Si IV, Si III for the different SED models described in  Figure 3.
 Solid lines are results from  CLOUDY.
Dotted lines are the observed values. The CLOUDY predictions for  O VI and N V
are out of the range for the SED model (c).  O VI is out of the range for the 
SED model (b).

Figure 7.---Spectral regions including the CO A-X 0-0, 1-0, 2-0, 3-0, 4-0 absorption
bands from the z = 2.8108 and 2.8132 components  in the rest
frame of the z = 2.8108 absorber toward PKS 0528-250. The positions of the 
absorption bands are indicated by tick marks. The lowest right panel is a
composite spectrum obtained by combining the 2-0 band with the
 0-0, 1-0, 3-0, 4-0 bands which are
shifted to the wavelength of the 2-0 band.

\newpage

\rm
\renewcommand{\baselinestretch}{1.15}
\scriptsize

\begin{flushleft}

~~~Table 1. The identifications of absorption lines  of QSO 0528-250

\end{flushleft}

\begin{tabular}{cccccclc}  \hline
No. & $\lambda_{obs}(\AA)$ & $\sigma(\lambda)$ & W$_{obs}$ & $\sigma(W)$ & SL& ID & z$_{abs}$\\ 
\hline
 &&&&&&\\
1 &	5221.93 & 0.12 & 0.392 & 0.040 & 9.9&	NiII 1370     &	2.8113\\
2  &     5224.56 & 0.22 & 0.235 & 0.036 & 6.6&  NiII 1370     & 2.8132\\
3 &	5247.71	& 0.07 & 1.875 & 0.049 & 37.9& AlII 1671     & 2.1409\\
(4)& 	5273.53 &0.31& 0.091&0.030&3.1&?&\\
(5) &	5293.72	& 0.44 & 0.11 &	0.04 &3.1&	CaI 2722      &	0.9444\\
6 &	5312.28 & 0.06 & 3.792 & 0.054 & 70.6& SiIV 1393     &	2.8115\\
7 &	5346.69	& 0.09 & 2.850 & 0.057& 50.2 &	SiIV 1402     &	2.8115\\
(8) &	5370.41	& 0.41 & 0.128 & 0.038 & 3.4&	NiII 1709     &	2.1413\\
9 &	5436.42	& 0.19 & 0.688 &0.049 & 14.0 &	MgII 2796     &	0.9441\\
10 &	5450.20	& 0.11 & 0.476 &0.036 &	13.1& MgII 2803     &	0.9440\\
11 &	5470.03	& 0.44 & 0.223 & 0.046 & 4.8& NiII 1741     &	2.1409\\
12 &	5477.68	& 0.09 & 0.892 &0.041 &	21.7 & CIV 1548      &	2.5381\\
13 &	5486.93	& 0.16 & 0.516 & 0.042 & 12.3& CIV 1550      &	2.5382\\
(14) &	5502.70	& 0.35 & 0.107 & 0.033 & 3.2&	NiII 1751     & 2.1410\\
(15) &	5519.40	& 0.63 & 0.123 &0.039 &	3.1& CO 1447(?)   &	2.8136\\
  &&&&&& Co II 1448(?)& 2.8117\\
16 &	5545.07	& 0.27 & 0.247 & 0.039 & 6.3&	NiII 1454     &	2.8114\\
17 &	5595.11	& 0.53 & 0.145 &0.041 & 3.5& NiII 1467     & 2.8120\\
18  &	5607.70 & 0.26 & 0.102 &0.029 &	3.5&Si II1526  & 2.6731\\
19 &	5614.02	& 0.42 & 0.153 &0.038 & 4.1& CIV 1548(?)    & 2.6262\\
(20) &    5622.57 & 0.34 & 0.107 & 0.032& 3.4& CIV 1550(?)   & 2.6256\\
21 &	5678.62	& 0.13 & 0.391 & 0.035 & 11.1 &Si II 1808    &	2.1408\\
22   &	5683.49	& 0.14 & 0.201 &0.030 &6.6  &	C IV  &	2.6710\\
23 &	5687.23	& 0.15 & 0.665 &0.047 &14.3&	CIV 1548      &	2.6734\\
24 &	5695.89	& 0.18 & 0.360 &0.040 & 9.1&	CIV 1550      &	2.6729\\
25 &	5818.72	& 0.02 & 5.008 &0.034 &	147.4& SiII 1526     &	2.8113\\
  &     5823.09 & 0.03 & 3.274 & 0.033 & 98.4& SiII 1526, AlIII 1855     & 2.8142\\
26 &	5849.26	& 0.18 & 0.385 &0.038 &	10.2& AlIII 1863    &	2.1401\\
27 &	5857.48	& 0.28 & 0.121 &0.030 &	4.0& ?   &	\\
(28)&  	5887.47 & 0.21 & 0.079 & 0.025& 3.1& CO 1544 (?)&  2.8122\\
29 &    5892.08 & 0.14 & 0.32 & 0.04 & 8.1& telluric line &       \\
30 & 	5900.16	& 0.04 & 4.889 &0.051 & 95.5&	CIV 1548      &	2.8110\\
31 &	5909.85	& 0.04 & 3.721 &0.049 &	76.5& CIV 1550      &	2.8110\\
32 &	5947.05	& 0.29 & 0.145 &0.037 & 3.9&	(?)      & 	\\
33&	6000.53&0.32&0.144&0.039&3.7& Co 1574(?)& 2.8110\\
&&&&&&&\\
\hline

\end{tabular}

\newpage
\scriptsize

\begin{flushleft}

Table 2. Line Strengths and Column Densities $^{a}$ 

\end{flushleft}

\begin{tabular}{lllccccc}  \hline \hline
~ & ~ & ~ & \multicolumn{3}{c}{z=2.8108} & \multicolumn{2} {c}{z=2.8132}\\ 
\cline{4-6}
\cline{7-8}

Species & $\lambda$ & f$^d$ & W$_{\lambda}(\AA)$ & N(cm$^{-2}$)$_{b=60 
km~s^{-1}}$ &N(cm$^{-2}$)$_{b=80 km~s^{-1}}$& W$_{\lambda}(\AA)$ & 
N(cm$^{-2}$)$_{b=30 km~s^{-1}}$ \\
\hline
Al II (A$_1$+A$_2$)$^b$& 1670.788& 1.88&2.100&$4.4\times 
10^{14}$&&2.100&$4.4\times 10^{14}$\\
C I & 1560.310 & 0.0822&  & $\le 6.0\times 10^{12}$$^e$ && $<$0.032& 
$<$1.8$\times 10^{13}$ \\
C II (A$_1$+A$_2$+A$^*_2$)$^b$ & 1334.532 & 0.118 & 2.29 & 1.7$\times 10^{17}$ 
&& 2.29 & 1.7$\times 10^{17}$\\
C II$^*$& 1335.702 & 0.118&  & && 0.312& 3.6$\times 10^{14}$\\
C III (A$_1$+A$_2$)$^b$& 977.020& 0.768 & 1.89& 7.6$\times 10^{16}$&& 1.89& 
7.6$\times 10^{16}$\\
C IV& 1548.202& 0.194&1.28& 5.6$\times 10^{15}$& 1.3$\times 10^{15}$&& \\
C IV&1550.774&0.0970& 0.976&2.0$\times 10^{15}$& 1.1$\times 10^{15}$&&\\
Si I$^c$& 1845.520& 0.229& $<$0.105&$<$1.6$\times 
10^{13}$&&$<0.105$&$<$1.6$\times 10^{13}$\\
Si II& 1808.013& 0.00208& 0.397&8.5$\times 10^{15}$&&0.094&1.7$\times 10^{15}$\\
Si II$^*$$^c$& 1533.431 & 0.13 & $<$0.031& $<$1.2$\times 10^{13}$& && \\
Si III& 1206.500& 1.66& 1.14&1.9$\times 10^{15}$&3.6$\times 10^{14}$ && \\
Si IV&1393.755& 0.528& 0.995& 7.9$\times 10^{14}$&3.0$\times 10^{14}$&&\\
Si IV&1402.770& 0.262& 0.748& 4.6$\times 10^{14}$&3.0$\times 10^{14}$&&\\
S I$^c$&1425.030&0.181&$<$0.031&$<$9.5$\times 10^{12}$&&$<$0.031&$<$9.5$\times 
10^{12}$\\
S II&1250.583 & 0.00535 & 0.157 & 2.7$\times 10^{15}$&& 0.131 & 2.6$\times 
10^{15}$\\
 &1253.808 & 0.0107 & 0.315 & ... && 0.236 & ...\\
S III (A$_1$+A$_2$)$^b$&1012.504 & 0.0355 & 0.622 & 3.2$\times 10^{15}$& & 
0.622 & 3.2$\times 10^{15}$\\
O I&1302.169 & 0.0486 & 1.36 & 2.1$\times 10^{17}$ && 0.70 & 1.1$\times 
10^{17}$\\
O VI & 1031.927 & 0.13 & 0.97 & 4.0$\times 10^{16}$&5.2$\times 10^{15}$&&\\
N V& 1238.821& 0.157&0.247& $1.5\times 10^{14}$& $1.4\times 10^{14}$&&\\
Fe II&2249.879 & 0.0018 & 0.160 & 1.8$\times 10^{15}$& & 0.046 & 3.6$\times 
10^{14}$\\
&2260.781& 0.0028& 0.260& ...&& 0.046& ...\\
&2374.461&0.0395&1.201&...&&0.342&...\\
Ni II& 1741.549&0.068& 0.116& 6.9$\times 10^{13}$&&0.072& 3.8$\times 10^{13}$\\
 & 1709.600& 0.047&0.101& ...&&0.056&...\\
 & 1370.132 & 0.100 & 0.103 & ... && 0.062 & ...\\
 & 1454.842 & 0.0515 & 0.065 & ... && 0.027 & ...\\
Ni II (A$_1$+A$_2$)$^b$ & 1467.756 & 0.0149 & 0.038 & 1.3$\times 10^{14}$&& 
0.038& 1.3$\times 10^{14}$\\
Co II (?)& 1448.011& 0.04516& 0.032& 3.9$\times 10^{13}$&&&\\
CO A-X (?)& 1447.359 & 0.0360 &  & && 0.032 & 6.4$\times 10^{13}$\\
CO (composite)$^c$&&&$<$0.017&$< 6.1\times 10^{12}$&&&\\
H$_2$ &&&& $2.6\times 10^{16}$&&&\\
\hline
\end{tabular} 

\medskip
$^a$ Rest frame equivalent widths,  obtained from Morton et al. (1980), Chen \& Morton
(1984), Foltz et al. (1988), Meyer et al. (1989),  Songaila \& Cowie (1995)
 and our observations.

$^b$ We adopt b = 100 km s$^{-1}$ when the absorption lines  from the
z = 2.8108 (A$_1$) and the z = 2.8132 (A$_2$) components  are blended. 

$^c$ The upper limits are calculated for 3 $\sigma$ deviations.

$^d$ Oscillator strengths, f, are from Morton (1991) except that  for Si II \lam\ 1808 is from Bergeson \& Lawler (1993) and Si 
II$^*$ 1533 is from Hibbert et al. (1992). 

$^e$  Songaila (1995, private communication).

\newpage

\begin{figure}
\plotone{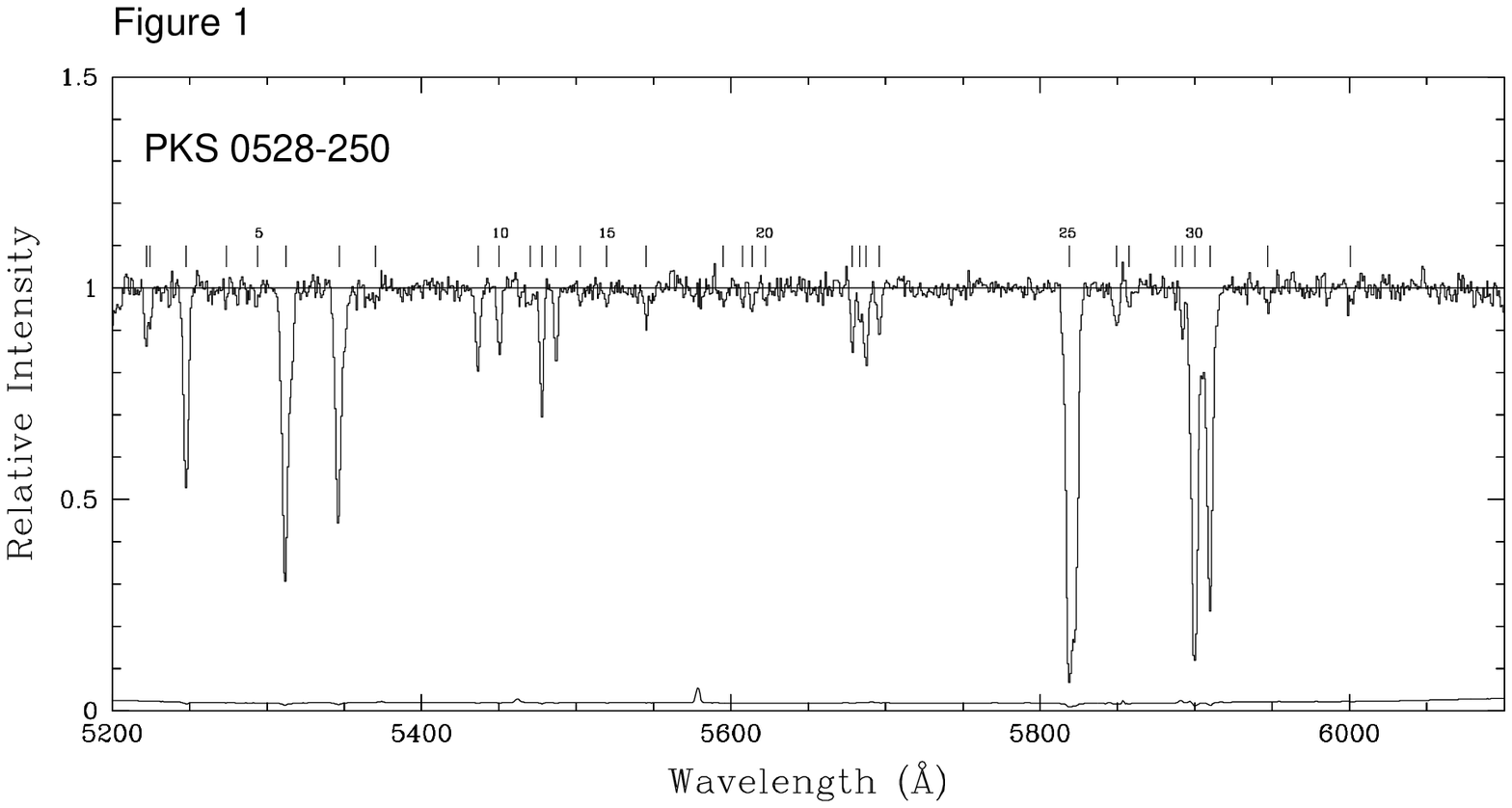}
\end{figure}

\newpage

\begin{figure}
\plotone{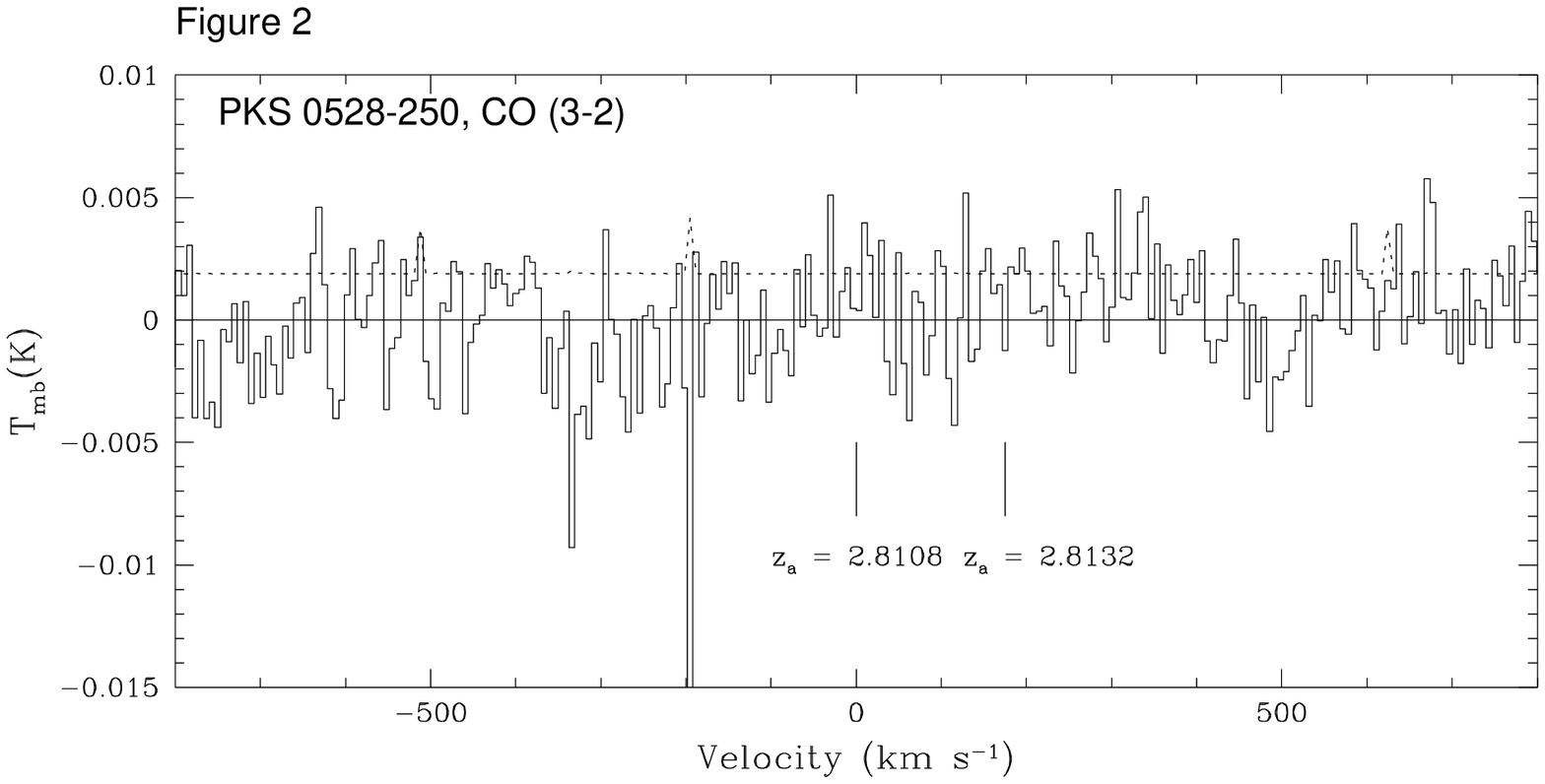}
\end{figure}

\newpage

\begin{figure}
\plotone{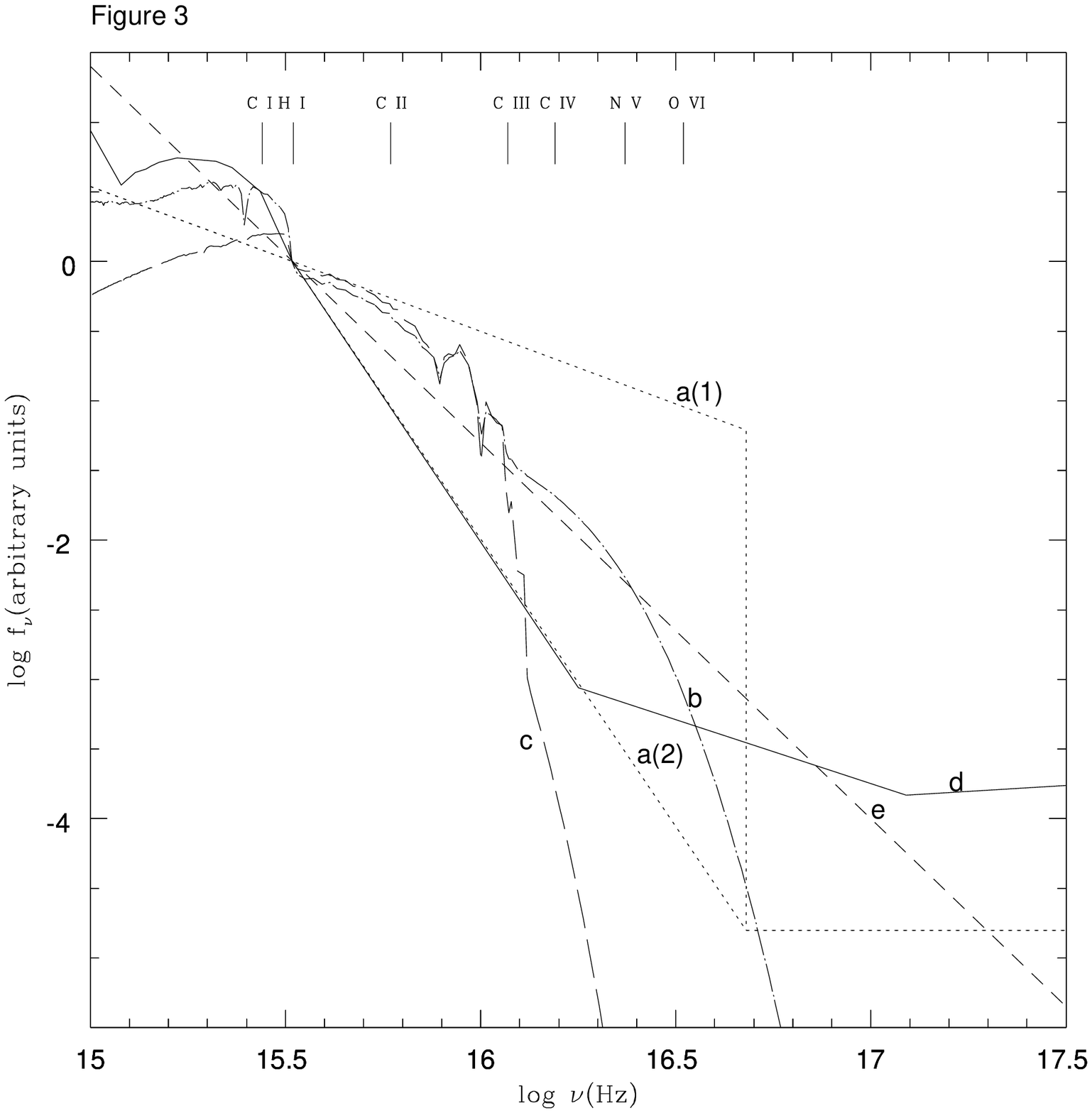}
\end{figure}

\newpage

\begin{figure}
\plotone{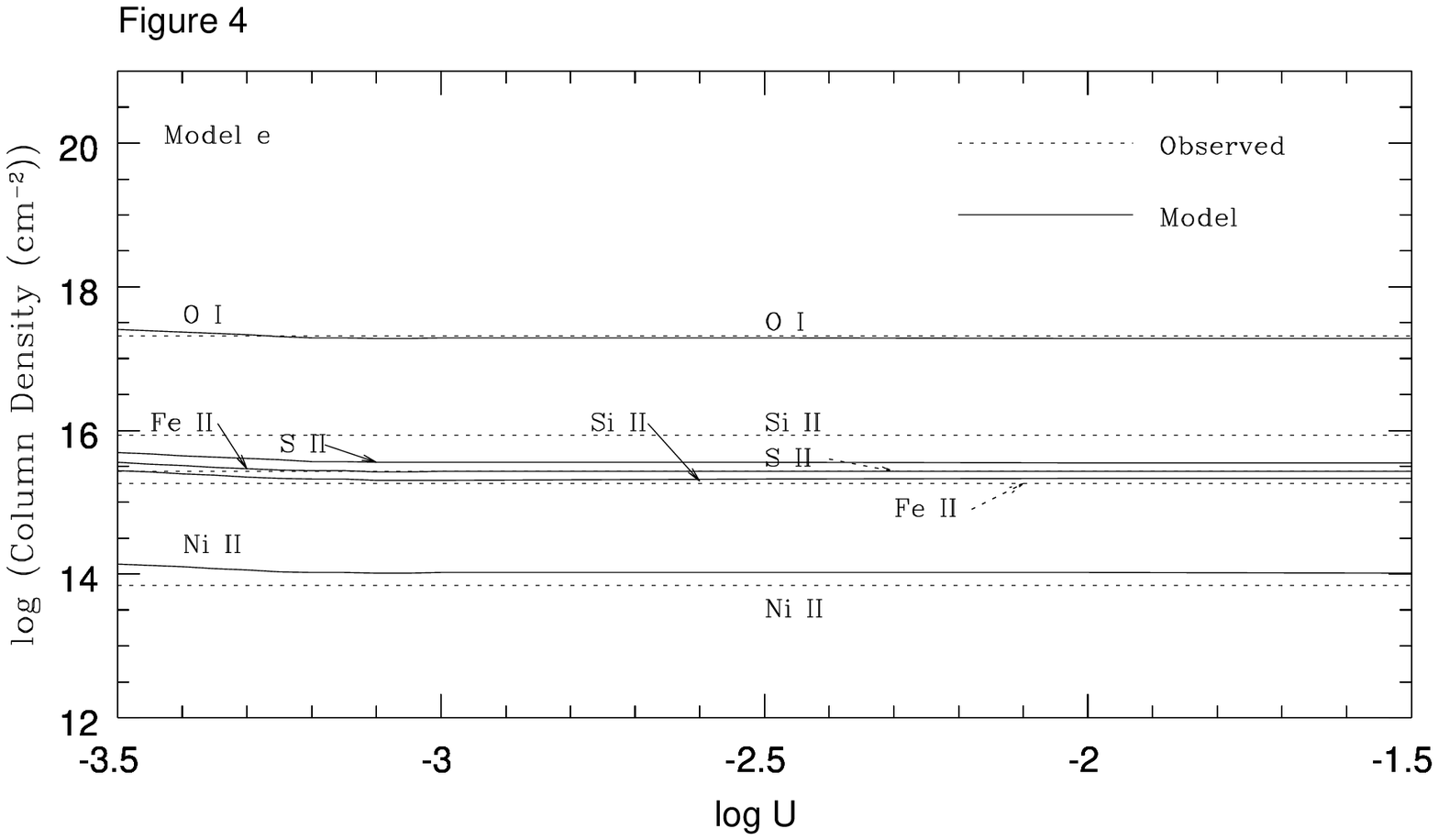}
\end{figure}

\newpage

\begin{figure}
\plotone{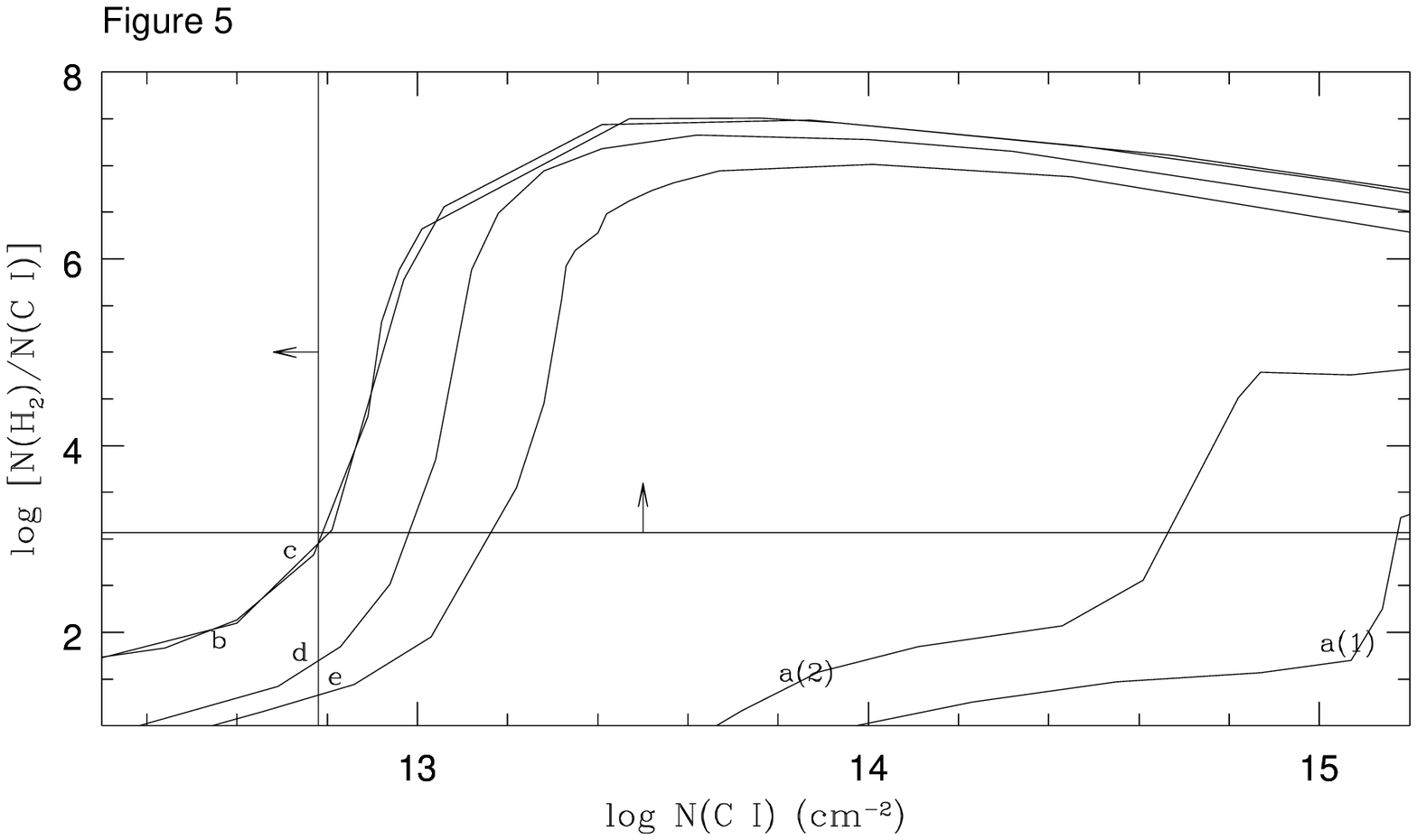}
\end{figure}

\newpage

\begin{figure}
\plotone{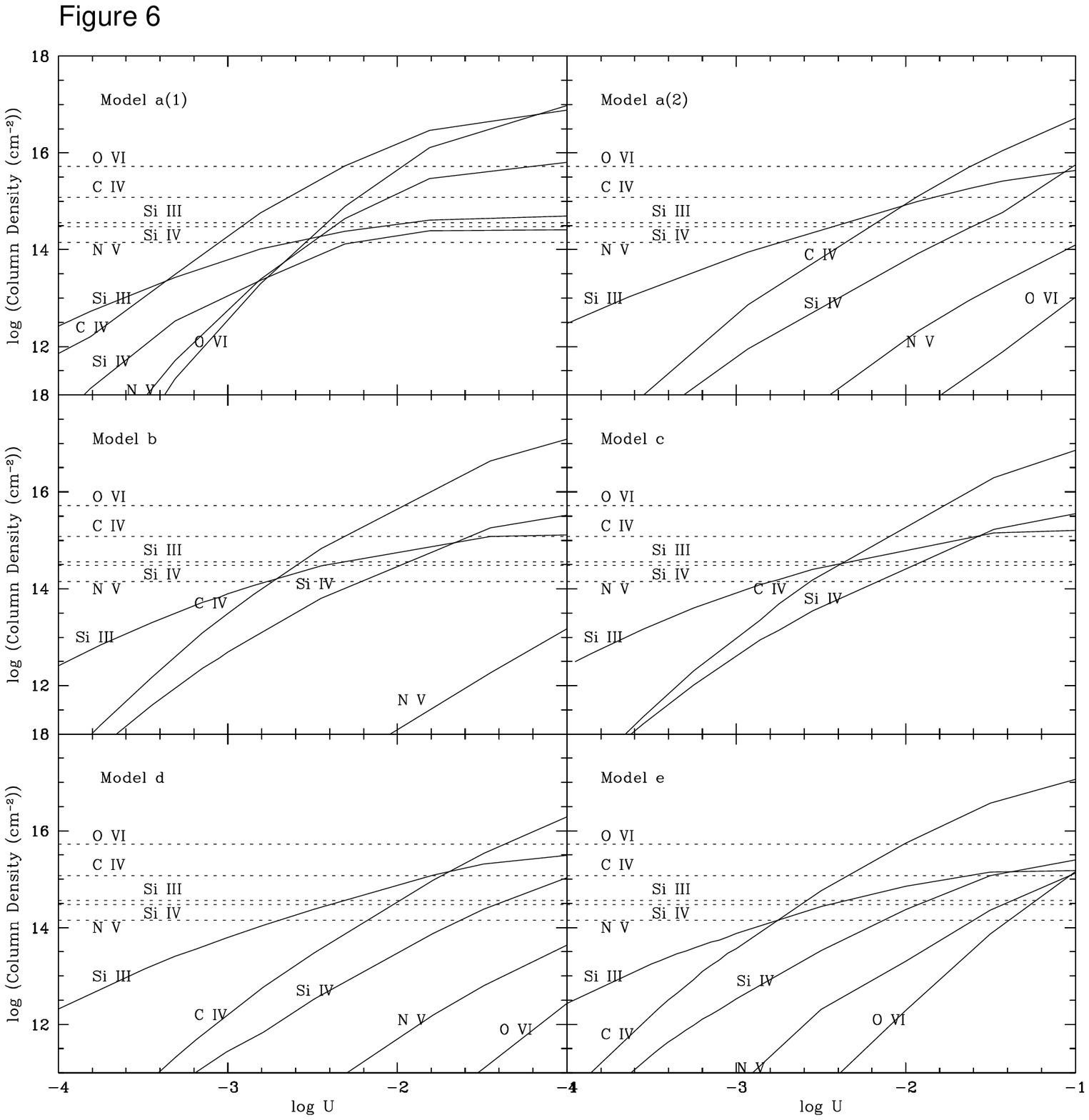}
\end{figure}

\newpage

\begin{figure}
\plotone{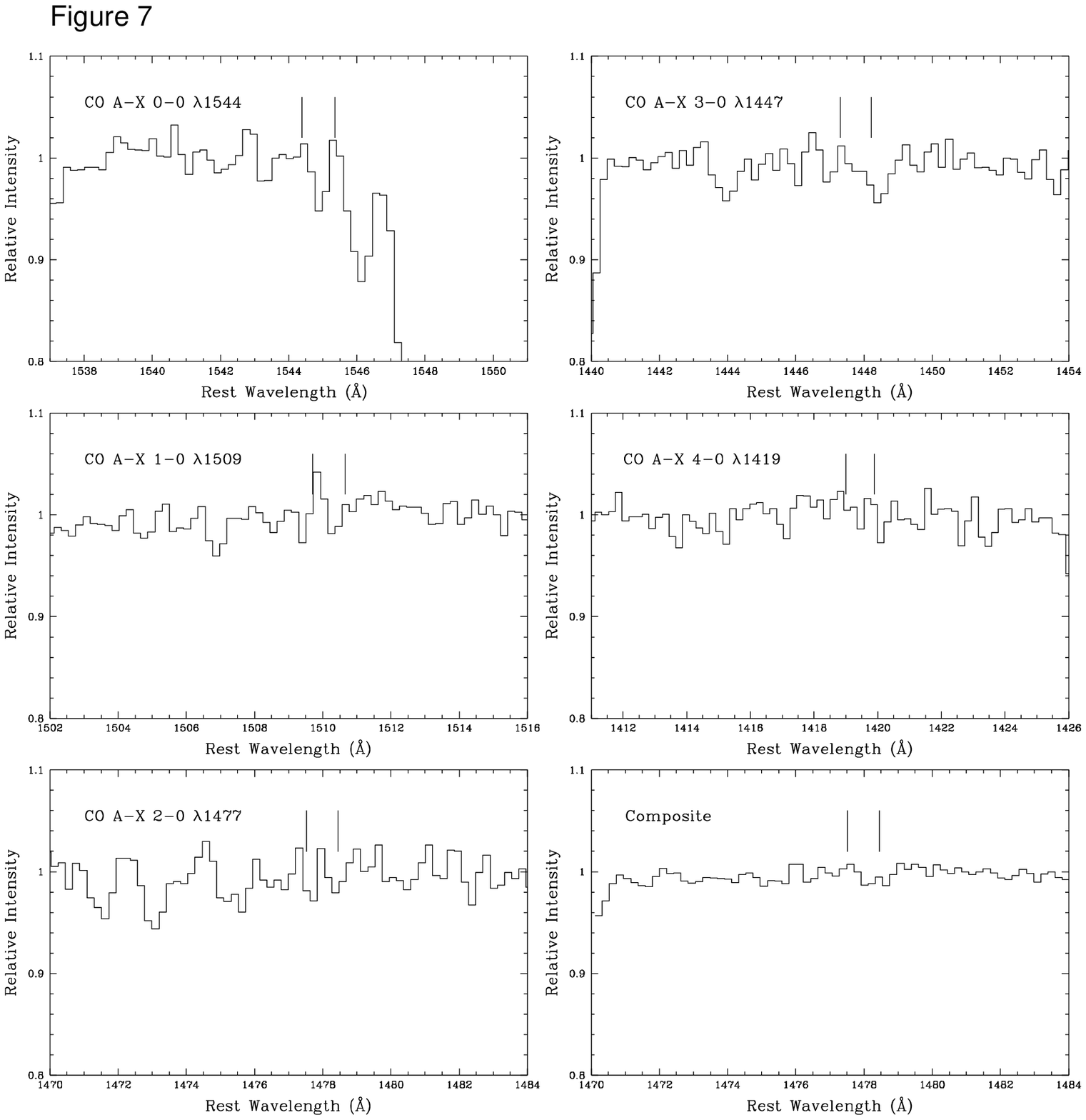}
\end{figure}

\end{document}